\documentclass[preprint2]{aastex}

\shorttitle{Universal Mass Function}
\shortauthors{Binggeli \& Hascher}

\begin{document}

\title{Is There a Universal Mass Function?}

\author{Bruno Binggeli and Tatjana Hascher
}
\affil{Department of Physics and Astronomy, University of Basel, 
Venusstrasse 7, CH-4102 Binningen, Switzerland}



\begin{abstract}
Following an old idea of Fritz Zwicky, we make an attempt 
to establish a ``universal'' mass function
for astronomical objects on all scales.
The object classes considered are: solar system planets and small bodies, 
exoplanets,
brown dwarfs, stars and stellar remnants, open and globular clusters, 
molecular clouds, galaxies, groups and clusters of galaxies. For comparison
we also include CDM halos taken from numerical 
simulations. We show that 
the mass functions of individual object classes, when properly
normalized, can indeed be concatenated
to build a surprisingly continuous mass function of the universe,
from $M \approx 10^{-20} {\rm M}_{\sun}$ (sub-kilometer size asteroids)
up to $M \approx 10^{16} {\rm M}_{\sun}$ (rich clusters of galaxies),
covering 36 orders of magnitude in mass.
Most individual mass functions roughly follow
a power law of the form $\phi(M) \propto M^{-2}$. A notable exception are
planets and small bodies which seem to obey a flatter distribution.
CDM halos from
high-resolution numerical simulations show a very similar relation, 
again of ``universal slope'' $-2$, from clusters of galaxies all the way 
down to the planetary mass scale. On the scale of stars and star clusters
this is a remarkable 
coincidence, as the formation processes involved are thought to be
totally different
(bottom-up gravitational clustering of DM halos 
versus top-down gravoturbulent fragmentation of gas clouds). 
\end{abstract}



\section{INTRODUCTION}
There is a great variety of astronomical objects in the universe --
by increasing average size or mass: asteroids, planets,
brown dwarfs, stars and stellar remnants, binary and multiple stars,
star clusters, molecular clouds, galaxies, groups and clusters of galaxies.
Each of these classes of objects follows a certain (but generally
not well known) distribution function in size, luminosity, or mass.
 The most fundamental characteristic property of all these distribution 
functions is that smaller or less massive 
things of a given kind are more abundant than larger or more massive ones. 
Presumably the same would hold not only within a certain class of objects,
but also if one were to glue together
the individual distribution functions to build
a ``universal'' distribution function, as there would be, {\em per unit volume
of space}, more asteroids than planets, more planets than stars, more stars
than galaxies, etc. 

In this paper we are doing just that: putting together all known distribution
functions for different object classes to see whether there is such a
thing like a ``universal distribution function''. -- ``Universality'' has a 
twofold meaning here: We can ask (1) whether the all-object distribution
function is sufficiently {\em continuous}\/ to render the notion of 
universality useful in the first place,
and (2) if yes, whether the {\em slope}\/ of the individual 
distribution functions and that of the grand total is, to a certain degree,
universally the same. For the present study we choose 
{\em mass}\/ as the independent parameter, as mass is probably 
the most fundamental -- albeit difficult to determine -- property of an 
astronomical object (luminosity is inappropriate because
there are dark objects). We show that by proper normalization we can indeed
establish a continuous mass function of the universe from $10^{-20} 
{\rm M}_{\sun}$
(100 meter-size asteroids) to at least $10^{16} {\rm M}_{\sun}$ 
(the most massive clusters of galaxies), 
and that there is surprising similarity in the slope of the mass function 
over a large range of mass.

The question of a universal distribution function of astronomical
objects has apparently been of interest only to Fritz Zwicky. 
In his 1942 Physical Review paper entitled
``On the large scale distribution of matter in the universe'',
Zwicky argued on purely theoretical grounds that the luminosity function (LF)
of galaxies had to be essentially exponential, i.e.~ever fainter galaxies 
being increasingly more numerous, just as observed today (see below) -- but in
contradiction with the data available then, which showed a more Gaussian-type
LF for galaxies. Zwicky's vision was that galaxies constitute an ensemble of
``particles'' in a stationary
state of statistical equilibrium. Frequent close encounters 
would dissolve larger aggregates of galaxies on the one hand, and 
accumulate new ones on the other hand, resulting in a Boltzman-type energy
distribution for galaxies and groups and clusters of galaxies
{\em as a whole}. Today, we know that the temperature and density of the
``galaxy gas'' is much too small, 
and thus the time-scale of galaxy interactions 
much too long, for a statistical equilibrium to be established or 
maintained.
But Zwicky was perfectly right not only in his prediction of an exponential LF
(which he himself set out to prove, see Zwicky 1957, p.~220ff), but also in his
view that galaxian systems, from dwarf galaxies up to the richest clusters
of galaxies, follow one, continuous distribution function (which was  
observationally established for the first time 
by Bahcall 1979, see also below). This unity of objects on the
scale of galaxies and systems of galaxies is now understood in terms
of a general, cosmological ``bottom-up scenario'' centred around the
hierarchical, gravitational clustering of dark halos in an expanding universe
having formed from primordial density fluctuations,
as, e.g., described by the ``Press-Schechter formalism''
(Press \& Schechter 1974). 

While Zwicky considered stars in his 1942 paper as well,
he was well aware that the mass distribution
of stars must be shaped by totally different physical processes.  
Put in simple terms, the relation between stars and (small) 
galaxies is plausibly
one of gas fragmentation (from large to small), rather than clustering (from
small to large). So why should there be a universal mass function, if
there can hardly be universal (global) 
physics behind the frequency distribution
of such things like asteroids, stars, and galaxy clusters?
This must be the reason why the question of a universal mass distribution
(beyond the scale of galaxies and systems of galaxies)
has never been taken up since Zwicky; it simply does not seem to make much
sense from a physical point of view. 

Conversely, the -- surprising -- degree of 
universality in the mass distribution of astronomical objects 
found in the present paper does not entitle us to conclude that there indeed
is overarching physics at work, though this 
cannot be excluded either. While we will, in the discussion section,
briefly indulge in speculations
on what it all could mean, we emphasize that our primary goal is
not to find a sort of cosmological principle that governs the mass distribution
function of the universe, but to provide a valuable {\em piece of cosmography}.
A good knowledge of the frequency distribution of things in the universe
is an end in itself.  

There are conceptual difficulties with the definition of an
astronomical object. What exactly is an astronomical object, i.e.~what 
kind of objects should we include in such a study? The most simple
requirement would be that such an object is gravitationally bound.
However, sub-kilometer asteroids (let alone dust particles,
if we were to push the distribution function to the extremes) are
hold together by non-gravitational forces. Also, on large scales, where
gravity does dominate, we wish, for the sake of continuity, to include 
molecular clouds which are in general unbound but are {\em identifiable 
entities}\/ none the less. 

Each class of objects has its own specific technical
difficulties in the establishment of a distribution function. There
is a rich history of attempts, e.g., to determine the stellar 
LF (and thus the initial mass function, IMF), or the LF and mass-to-light (M/L)
ratio of galaxies. The present paper is not meant to be
a meta-study where all these attempts
for the different object classes are critically assessed. Rather, we will
take ``reasonable'' distribution functions from the recent literature,
with no consideration of the uncertainties,
and put them together to construct a {\em first-guess universal 
mass function}.
These distribution functions will partially overlap in mass and partially be 
separated by gaps in mass regions where we lack objects or data.  

Another principal difficulty of such a study is the normalization.
Everything has to be normalized to the same volume of space, but our objects 
are sampled in vastly different volumes: While
clusters of galaxies are seen over a significant part of the observable
universe, and the galaxian LF, though sampled in a fairly local volume, 
can still be regarded as universal, the stellar LF
is simply that of the solar neighbourhood, 
and the mass function of planets and small
bodies is entirely drawn from our solar system (never mind the exponentially 
growing number of -- nearby -- exoplanets). So the smaller the
scale the larger and more uncertain the extrapolation to a universal mean.  
This kind of uncertainty will not soon go away. Rather, we have to rely,
as all of cosmology does, on the Cosmological Principle, i.e. that 
we are living
in a typical, average environment -- on all scales. All our mass functions will
in the end be normalized to a volume of 1 Mpc$^3$ and given in units of solar
masses ([$\phi(M)] = {\rm Mpc}^{-3} {\rm M_{\sun}}^{-1}$).

The paper is organized as follows. In section 2 we present the 
individual mass functions for the different object classes, starting with 
galaxies and going up first to supergalactic, then going down to subgalactic
structures. These functions are put together to a universal mass function
in section 3, where the 
mass distribution of dark matter halos from theory is added for comparison.
Finally, in section 4 we discuss possible meanings of the 
universality found and give our conclusions. A Hubble constant of 
70 km s$^{-1} {\rm Mpc}^{-1}$
is used throughout.

\section{OBSERVED MASS FUNCTIONS FOR DIFFERENT CLASSES OF OBJECTS}

\subsection{Galaxies}
While the galaxian LF is fairly well known, except at the very faintest end,
the mass function (hereafter MF) of galaxies is very hard to
determine, as galaxies are dominated by non-baryonic dark matter (DM)
of a principally unknown nature. The galaxian MF is reconstructed
from individual or statistical mass-to-light-ratio determinations
(or rather estimates!) that have 
to be convolved with the luminosity distribution 
which is usually taken to be a Schechter (1976) function of the form\\
\begin{equation}
\phi(L) dL = \phi^*(L/L^*)^\alpha \exp(- L/L^*) dL\,\,\,.
\end{equation} 
Vale \& Ostriker (2004) give the following mass-to-light 
function for galaxies:\\
\begin{equation}
L(M) = A \frac{(M/M')^b}{[c + (M/M')^{d\,k}]^{1/k}}
\end{equation}
with the best-fitting parameters $A = 5.7 \times 10^9$, $M' = 10^{11}$, 
$b$ = 4, $c$ = 0.57, $d$ = 3.72, and $k$ = 0.23. At the low-mass end this
gives $L \propto M^4$, at the high-mass end $L \propto M^{0.3}$.
A total galaxy MF is then obtained through 
\begin{equation}
\phi(M) = \frac{dL(M)}{dM} \phi(L(M))
\end{equation} 
by inserting (2) and the Schechter function (1) with parameters
taken from the 2dF galaxy LF (Norberg et al.~2002), as used also by
Vale \& Ostriker (2004): $\phi^* = 5.52 \times 10^{-3} 
{\rm Mpc}^{-3} {\rm L}_{\sun}^{-1}$,
$L^* = 2.31 \times 10^{10} {\rm L}_{\sun}$ 
(in the $B$ band), and $\alpha = -1.21$.
However, we use this Vale \& Ostriker galaxy MF only for galaxy masses 
smaller than $10^{11} {\rm M}_{\sun}$, as the high-mass end (with its shallow
slope, see above) would produce too many cD-type galaxies (as Vale and Ostriker
note, the cD halos that went into their mass-to-light function are rather
hosting the whole cluster than the cD alone).  
A more appropriate MF at the high-mass end may be provided by van den
Bosch et al.~(2005) who work with the statistical distribution of satellite
galaxies. Their mass-to-light function is given by 
\begin{equation}
L(M) = \frac{2 M}{(M/L)_0 [(M/M_1)^{-\gamma1} + (M/M_2)^{\gamma2}]}
\end{equation}
with the best-fitting parameters $(M/L)_0$ = 115, log $M_1$ = 10.76,
log $M_2$ = 12.15, $\gamma1$ = 2.92 and $\gamma2$ = 0.29 (for a 
$\Lambda$CDM clustering normalization of $\sigma_8$ = 0.9). This is
somewhat steeper at the high-mass end ($L \propto M^{0.7}$ versus $M^{0.3}$), 
giving a
total MF that is falling down more rapidly now with increasing mass, 
while it is roughly compatible with Vale \& Ostriker (2004) at the 
low-mass end. We therefore use the resulting van den Bosch et al. 
MF for galaxies with masses larger than $10^{11} {\rm M}_{\sun}$, keeping
however the Vale \& Ostriker MF for smaller masses, with continuity 
at $M = 10^{11} {\rm M}_{\sun}$. This {\em total}\/ (DM-dominated) 
galaxy MF, as a combination of the two functions for larger and smaller masses
as just described, is shown in Fig.~1. The universal normalization 
(per Mpc$^3$) is of course provided by the adopted normalized LF.

For comparison we also show the baryonic MF of galaxies in Fig.~1
(stars, stellar remnants, atomic and molecular gas), taken
from Read \& Trentham (2005). It is
based on a host of NIR broad-band as well as radio observations and is
given in the form of a Schechter mass function
\begin{equation}
\phi(M) dM = \phi^*(M/M^*)^\alpha \exp(- M/M^*) dM
\end{equation} 
with parameters $\phi^* = 2.5 \times 10^{-14}  
{\rm Mpc}^{-3} {\rm M_{\sun}}^{-1}$, $M^* = 
1.31 \times 10^{11} {\rm M_{\sun}}$,
and $\alpha = -1.21$. The baryonic galaxy mass is indeed dominated by stars;
the Schechter $\alpha$ is the same
(see Read \& Trentham 2005 for a detailed account of the contribution by the
gas). We show the galaxy MFs down to a mass of $10^{6} {\rm M_{\sun}}$; this
is of course a bold extrapolation; the faint end of the LF, let alone the MF
of galaxies, is not well constrained.

\begin{figure}[tb]
\plotone{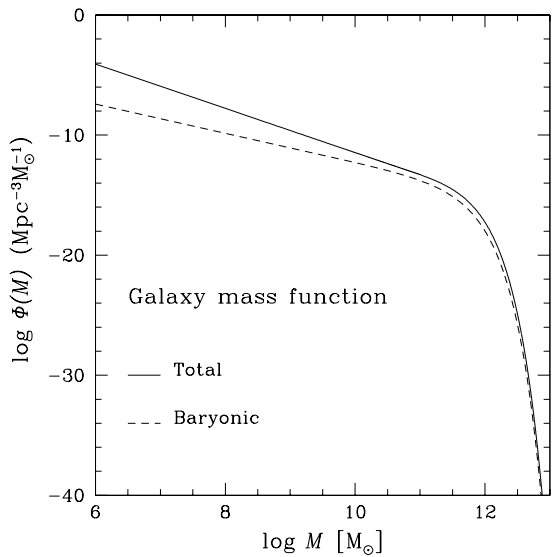}
\caption{Total and baryonic mass functions of galaxies.}
\end{figure}

\subsection{Groups and Clusters of Galaxies}
The mass of groups and clusters of galaxies is dominated by DM as well
and has to be estimated either from the kinematics of member galaxies
or (for clusters only) from the X-ray intracluster gas properties or the
gravitational lensing of background galaxies. There are several 
studies of the group and cluster MF in the literature
(e.g.~Bahcall \& Cen 1993, Biviano et al.~1993, Girardi et al.~1998).
For our purposes we take the MF of Girardi et 
al.~(1998) which is based on optical virial mass estimates of a complete
sample of 152 nearby
clusters. It was derived only for masses larger than a few 
$10^{14} {\rm M_{\sun}}$, but we extrapolate it down to $10^{12} 
{\rm M_{\sun}}$,
because it is compatible with the MF of Bahcall \& Cen (1993) which does cover
also the low-mass end by including small groups. The adopted MF for groups and
clusters, shown in Fig.~2, is a Schechter-type MF of the form (5), 
however fixing $\alpha = -1$ (flat end), with  
$\phi^* = 6.2 \times 10^{-20}  
{\rm Mpc}^{-3} {\rm M_{\sun}}^{-1}$, and $M^* = 
3.7 \times 10^{14} {\rm M_{\sun}}$.  

\begin{figure}[tb]
\plotone{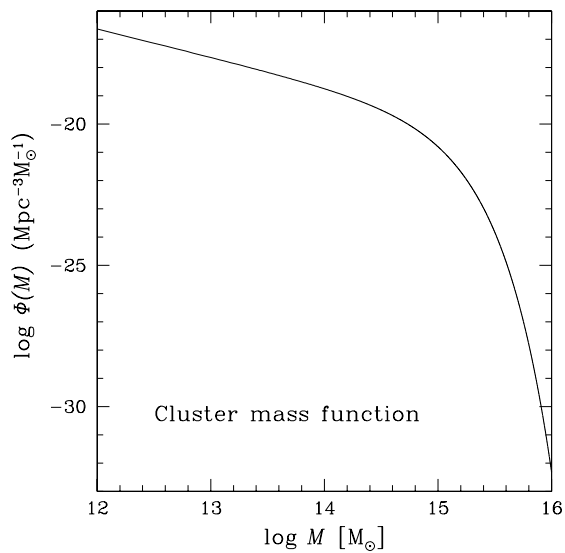}
\caption{The mass function of groups and clusters of galaxies }
\end{figure}

\subsection{Star Clusters and Molecular Clouds}
Next we consider subgalactic structures: star clusters and molecular clouds
which can be as massive as extreme dwarf galaxies but in contrast to these
do not contain dark matter.
\subsubsection{Globular clusters}
The observed LF of globular clusters is typically bell-shaped (Gaussian). 
However, the decline towards fainter luminosities (masses)
is probably an effect of dynamical evolution (dissolution of smaller clusters
in the galactic tidal field) and may not reflect the initial distribution
[for a different view, however, see Parmentier \& Gilmore 2007].
As suggested by Surdin (1979), Racine (1980), and Harris \& Pudritz (1994),
the initial mass distribution
of globular clusters can be modelled by a power law of the form
\begin{equation}
\frac{dN}{dM} \propto M^{-{\beta}}\,\,\,
\end{equation} 
or in our MF notation:
\begin{equation}
\phi(M) dM = A M^{-{\beta}} dM\,\,.
\end{equation}
Such a distribution is indeed observed for young clusters in
merging galaxies (e.g. Whitmore \& Schweizer 1995), and the high-mass tails
of the old globular cluster systems of nearby galaxies are still following
this form very closely. For globulars with masses larger than $10^{5} 
{\rm M_{\sun}}$, Harris \& Pudritz (1994) find a mean best-fitting 
$\beta = 1.7$, which we adopt for our globular cluster MF in the range
$10^{5}-10^{7} {\rm M_{\sun}}$. For lower masses we consider only open
clusters (see below).

How can this MF now be normalized to a universal density? This is done 
by the ``specific frequency''
\begin{equation}
S_N = 8.55 \times 10^7 \frac{N_{\rm T}}{L_V/{\rm L}_{\sun}}\,\,\,,
\end{equation}  
which is the number of globular clusters per ``piece'' of mother galaxy
of absolute $V$ magnitude $-15$. $N_{\rm T}$ is the estimated total number of
globular clusters. $S_{\rm N}$ depends on the morphological type of the
mother galaxy, but a reasonable mean value is $S_{\rm N} = 
3$ (see Harris 2001),
which we adopt. With a universal $V$ luminosity density, $\rho_{\rm L}$, 
we have
for the mean number of globular clusters per Mpc$^3$:
\begin{equation}
A_{\rm gc} \hspace{-2mm}\int\limits^{10^7 {\rm M}_{\sun}}_{10^5 
{\rm M}_{\sun}} \hspace{-2mm}M^{-1.7} dM = 
\frac{S_{\rm N} [\rho_{\rm L}/{\rm L}_{\sun}]}{8.55 \times 10^7}\,\,.
\end{equation}  
Norberg et al.~(2002) give a $B$ band $\rho_{\rm L} = 1.27 \times 
10^8 {\rm L}_{\sun}
{\rm Mpc}^{-3}$. As the mean $B-V$ colour of galaxies will not be far
from the solar value of 0.65, we assume the same $\rho_{\rm L}$ value in $V$.
The resulting normalization constant from (9) is then 
$A_{\rm gc} = 1.03 \times 10^4
{\rm Mpc}^{-3} {\rm M_{\sun}}^{-1}$. 

\subsubsection{Open Clusters}
As shown, e.g., by Elmegreen \& Efremov (1997) from their study
of hundreds of open clusters in the LMC, the LF of open clusters 
systematically depends on age. Using the same power law for the LF,
\begin{equation}
\phi(L) dL \propto L^{-{\beta}} dL\,\,,
\end{equation}
$\beta$ is smaller (i.e.~the LF more flattened) the younger the cluster.
Very young clusters (age less than $10^7$ years) have essentially a flat
distribution ($\beta \approx$ 1); somewhat older clusters up to an 
age of $10^8$ years show $\beta \approx$ 1.5, and for clusters older than  
$10^8$ years we finally have $\beta \approx$ 2. The flat LFs found with
young clusters is obviously due to the presence (and dominance, in terms of
light) of short-lived OB stars. Clearly, then, for the 
{\em mass}\/ distribution (7) of open clusters,
$\beta$ = 2 is a reasonable choice.
A normalization (fixing constant $A$ in (7) for open clusters) is achieved
by comparing the populations of open and globular clusters in the 
Milky Way galaxy. For open clusters we rely on van\,den\,Bergh \&
Lafontaine (1984) who, aside from finding $\beta \approx$ 2 as well,
give a mean surface density for open clusters in the solar neighbourhood
of $N_{\rm oc} \approx 30\,\,{\rm kpc}^2$ [see Piskunov et al.~(2006) and 
references therein for more recent open cluster population studies].
This is for clusters of absolute 
$V$ magnitude between $-2$ and $-10$, roughly corresponding to a mass range
of $10^3$ to $10^5$ solar masses. If we assume a more or less constant
cluster density in the galactic disk with 10 kpc radius, giving a total
disk surface of $\approx 300\,\,{\rm kpc}^2$, 
we expect a total number of galactic
open clusters of ca. 30 $\times$ 300 = 9000. As there are
about 150 globular clusters in the Galaxy, we have
\begin{equation}
\frac{A_{\rm oc} \int\limits^{10^5 
{\rm M}_{\sun}}_{10^3 {\rm M}_{\sun}} M^{-2} dM}
{A_{\rm gc} \int\limits^{10^7 
{\rm M}_{\sun}}_{10^5 {\rm M}_{\sun}} 
M^{-1.7} dM} = \frac{9000}{150}\,\,\,.
\end{equation}
With $A_{\rm gc}$ from above, this gives $A_{\rm oc} =  
1.89 \times 10^5 {\rm Mpc}^{-3} {\rm M_{\sun}}^{-1}$.

\subsubsection{Molecular Clouds}
Although molecular clouds are not gravitationally bound and are 
not even in a state of long-term equilibrium, being formed and dispersed 
on a time-scale shorter than the dynamical time-scale of the Galaxy,
they are objects in the sense that one can measure their sizes and masses
and thus determine a distribution function. Molecular clouds exhibit a
fractal structure, i.e.~smaller (and denser) clouds are embedded in larger
ones. The cumulative size distribution for substructures inside a 
fractal is given by  
\begin{equation}
N(>S) \propto S^{-D}\,\,,
\end{equation}
or in differential form
\begin{equation}
n(S) dS \propto S^{-(D+1)} dS\,\,,
\end{equation} 
where $D$ is the fractal dimension (Mandelbrot 1982; see 
Elmegreen \& Falgarone 1996, whose formalism we follow here). 
On the other hand, there is a size-mass relation which can be
assumed to be a power law
again:
\begin{equation}
M \propto S^{\kappa}\,\,.
\end{equation}
The size distribution can now be transformed into a mass distribution:
\begin{equation}
n(M) dM = n(S) \frac{dS}{dM} dM \propto M^{-\frac{D}{\kappa} - 1} dM\,\,.
\end{equation}
Observations show that $\kappa \approx D \approx 2.3$ (see 
Elmegreen \& Falgarone 1996). This is
consistent with another definition of the fractal dimension:
$M \propto S^D$ (Mandelbrot 1982), suggesting that the size and mass
distribution of molecular clouds is, in fact, the result of fractal gas
structure. The MF for molecular clouds
can therefore be written again as
\begin{equation}
\phi(M) dM = A_{\rm mc} M^{-2} dM\,\,.
\end{equation}

To fix the normalization constant $A_{\rm mc}$ we assume that the mean density
of molecular hydrogen in the universe is entirely made up of clouds with 
masses between $10^{-1} {\rm M}_{\sun}$ and $10^{7} {\rm M}_{\sun}$
(see Elmegreen \& Falgarone 1996 for the mass range):
\begin{equation}
\Omega_{\rm H2} \rho_{\rm crit} = A_{\rm mc} \hspace{-3mm}\int\limits_{
10^{-1} {\rm M}_{\sun}}^{10^{7} {\rm M}_{\sun}} \hspace{-3mm}\phi(M) M dM
= A_{\rm mc} \hspace{-3mm}\int\limits_{10^{-1} {\rm M}_{\sun}}^{10^{7} 
{\rm M}_{\sun}} \hspace{-2mm}M^{-1} dM\,\,.
\end{equation} 
With $\rho_{\rm crit} = 1.36 \times 10^{11} {\rm M}_{\sun} {\rm Mpc}^{-3}$
and $\Omega_{\rm H2} = 0.00026$ from Read \& Trentham (2005),
we get $A_{\rm mc} = 1.92 \times 10^6 {\rm Mpc}^{-3} {\rm M}_{\sun}^{-1}$.

The MFs we have derived so far are plotted together in Fig.~3.
We note, as has been noted before, that the MFs of molecular clouds and 
star clusters (and, in fact, of stars as well, see below) have very 
similar power laws. Such would be a natural outcome, if
stars and star clusters form(ed) by the fragmentation of turbulent 
cloud cores (e.g.~Elmegreen \& Efremov 1997).
   
\begin{figure}[tb]
\plotone{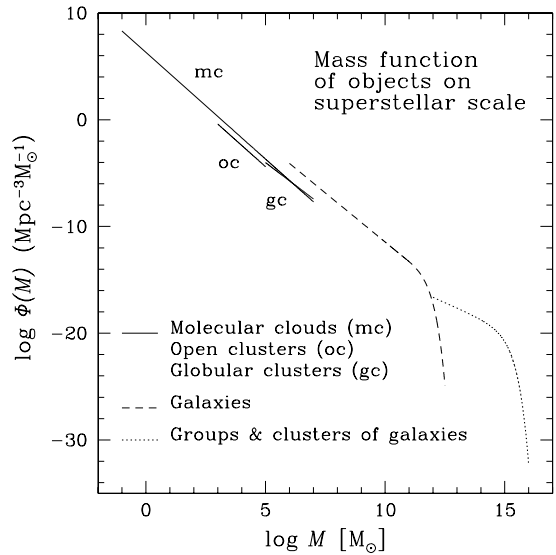}
\caption{The mass function of superstellar objects: molecular clouds,
open clusters, globular clusters, galaxies, and groups and clusters of 
galaxies.}
\end{figure}

\subsection{Stars and Stellar Remnants}
In the spectrum of cosmic objects we would next have to deal with 
binary and (the much less frequent) multiple stars before coming to 
single stars. However,
we are still lacking reliable binary statistics.
A very rough estimate of the MF of binary stars
is the following. Consider the extreme case that every star is in a system
of two stars of equal mass. Then the mass function of binaries per unit volume
would obviously be lowered in amplitude by at least 
a factor of two as compared to the MF of 
single stars, while otherwise being of similar shape and covering essentially 
the same mass range. This shows that in our first-guess
universal mass function, binary and multiple star systems can be neglected.

In the following we will determine the MF of a universally averaged, 
{\em evolved}, present-day
population of single stars {\em and stellar remnants}. Such a distribution
function has never been discussed in the literature, precisely because it
is interesting only in the present (new) context. 
   
\subsubsection{IMF and Final-to-Initial Mass Ratio}
The mass function of single stars at their birth, i.e. the initial 
mass function
(IMF), is an imortant datum for all of astronomy and is of course the
backbone of our treatment. The original form given by
Salpeter (1955):
\begin{equation}
\phi(M) dM \propto M^{-2.35} dM
\end{equation}
is still in use for stars of higher mass. For low-mass stars various 
modifications of the Salpeter function have been suggested (e.g., 
Kroupa 2001), 
but there is agreement that well below one solar mass the power-law index is 
around $-1.8$, rather than $-2.35$. For our purposes we adopt an index of
$-1.8$ for masses smaller than 0.5 M$_{\sun}$ (down to $M$ = 0.1 M$_{\sun}$), 
and the Salpeter index for stellar masses above that limit (up to 
$M$ = 120 M$_{\sun}$). 

A second, even less well-known ingredient for our calculation is the 
relation between the initial mass of a star and the final mass of its
remnant (white dwarf, neutron star, black hole). Stars with initial 
masses smaller than about 8 M$_{\sun}$ end up as white dwarfs; stars
with initial masses above this critical limit end up as neutron stars
or black holes. For the initial-mass range 1 to 7 
M$_{\sun}$ and solar metallicity (which we assume here for simplicity)
we adopt the final white dwarf masses given by Weidemann (2000, Table 2.1);
they range from 0.55 M$_{\sun}$ for a 1 M$_{\sun}$ star initially, to 1.02
M$_{\sun}$ for a 7 M$_{\sun}$ star initially. For subsolar masses, where little
is known about the initial-final mass ratio, we assume a simple
linear relation from ($M_{\rm i}/M_{\rm f}$) 
= (0.1, 0.1) to (1, 0.55). The exact 
relation is nearly irrelevant for our purposes because these stars 
do not evolve in a Hubble time, anyway.

For high-mass stars ($M_{\rm i} > 8$ M$_{\sun}$) the situation is very complex,
as the evolution and therefore also the mass
loss of these stars strongly depends on metallicity and other parameters,
such as rotation etc. (Woosley, Heger \& Weaver 2002; Heger et al.~2003).
We have constructed an initial-final mass function for these stars,
drawing on the information contained in the papers just mentioned, by extending
the low-mass relation continuously with 
a linear relation from ($M_{\rm i}/M_{\rm f}$) = (7, 1.02) up to 
(15, 1.4), and then
again linearly from (15, 1.4) all the way up to (120, 25). This is a very 
rough approximation (in lack of better knowledge) and we assume solar
metallicity (metal-poor stars would suffer less mass loss, resulting in
higher-mass remnants). 
The adopted initial-final mass function is shown in Fig.~4.

\begin{figure}[tb]
\plotone{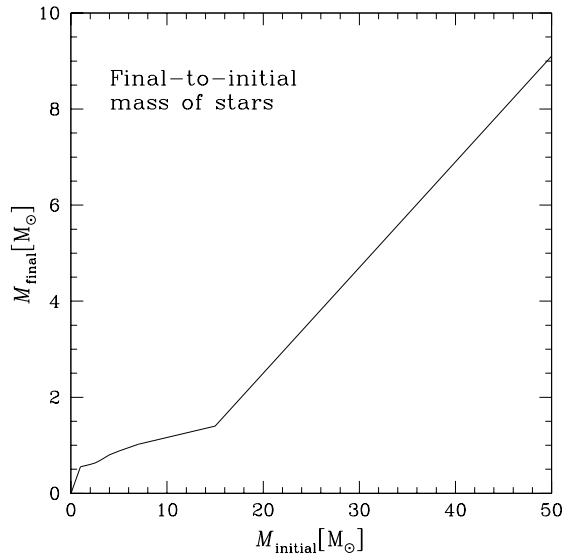}
\caption{The final mass of the stellar remnant as a function of the initial
mass of the star. Above 50 M$_{\sun}$ (initial mass) the linear relation
is extrapolated up to the highest masses considered.}
\end{figure}

\subsubsection{Stellar Lifetimes and Star Formation Histories}
Two more ingredients for our calculation of the MF of stars and stellar
remnants for an evolved stellar system are needed: the lifetime of a star
as a function of its initial mass, and the star formation rate as a function
of time (i.e.~the star formation history). The stellar lifetime (which is
essentially the time spent on the main sequence) is also a
function of metallicity; for simplicity we again assume solar metallicity.
A table with the stellar lifetimes for solar-metallicity stars with initial
mass between 0.8 and 120 M$_{\sun}$ is given, e.g., in the textbook of
Sparke \& Gallagher (2000); for stars with mass below 0.8 M$_{\sun}$ the
lifetime is larger than a Hubble time.
We have fitted a polynomial of second degree
to the tabulated pairs (log stellar mass, log lifetime) to get the
estimated lifetimes for stars of any (non-tabulated) mass.

Three types of star formation histories have been used as input: 
(1) an initial
burst of star formation, i.e.~all stars are born at the same instant,  
approximately applying to elliptical and S0 galaxies, (2) a constant 
star formation rate, simulating the case of spiral and irregular galaxies,
and (3) a ``cosmic'' star formation history. The latter is inferred from 
high-redshift observations and is usually given as star formation rate 
per unit volume as a function of redshift. We have adopted the 
log (SFR)--z 
relation of Madau, Pozzetti \& Dickinson (1998, Fig.~3) and turned it into a 
log (SFR)--t relation with
the help of a ``Cosmology Calculator''\footnote{
http://nedwww.ipac.caltech.edu/index.html Cosmology Calculator I}
using $H_0 = 71$ km s$^{-1}$ Mpc$^{-1}$, $\Omega_{\rm M} = 0.270$ and 
$\Omega_{\rm Vac} = 0.730$. The result is shown in Fig.~5.  

\begin{figure}[tb]
\plotone{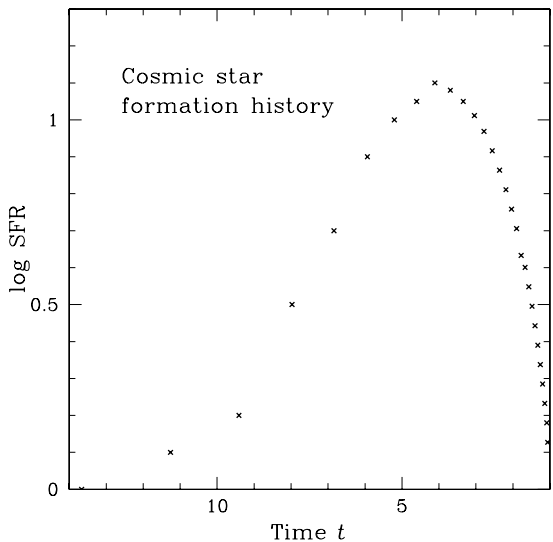}
\caption{The cosmic star formation history: the average 
star formation rate (SFR,
the mass of stars born per unit volume per unit time) of the universe
as a function of time (SFR in arbitrary units).}
\end{figure}

\subsubsection{Mass Function of an Evolved Stellar Population}
The present-day MF of stars and stellar remnants for an assumed
star formation history is now calculated in the following way. 
The whole mass range from 0.1 to 120 M$_{\sun}$ is divided into 
bins of width 0.1 M$_{\sun}$, giving 1\,200 mass points. 
The evolutionary timespan considered is 12 Gyr, i.e.~we assume
that the first stars were built around 12 Gyr ago, in accord
with the typical age of Milky Way globular clusters. Dividing this
timespan into intervals of 0.1 Myr gives 12\,000 timesteps. Consider
now that we start with a star formation event. The 
original mass function $\phi (M)$ is then simply the adopted IMF.
After every time step we compare the time elapsed since the birth
event with
the lifetime of a star of given mass. Stars with lifetimes smaller 
than the population age 
are instantly lost, i.e. taken away from the mass
function at their initial mass and added at a different
place of the mass function at their much 
smaller remnant (final) mass. In this way stars of subsequently 
smaller masses are lost from the population, leading to the 
well-known ``burning down'' of the IMF from high to low masses.
Non-coeval stellar populations (constant SFR or cosmic SFH) can 
easily be calculated as superposition of many little bursts 
distributed over the whole timespan (one burst for a given 
population fraction at every timestep). 

The resulting present-day
MFs of stars and stellar remnants for the three different
star formation histories considered are shown, along
with the IMF, in Figs.~6 and 7. The MFs are normalized to the
same integrated, {\em initial}\/ stellar mass as contained in 
the IMF; the normalization of the IMF itself is arbitrary.
The MF would in principle be unchanged, i.e.~remain identical
to the IMF, for masses somewhat 
smaller than 1 M$_{\sun}$. However, we note a little spike
on top of the IMF around $\log M = -0.2$, or $M \approx$ 0.6 M$_{\sun}$.
This is due to the ``graveyard'' of white dwarfs of initial mass
1 M$_{\sun}$ or smaller. The depletion at masses larger than about  
4 M$_{\sun}$ ($\log M > 0.6$) is
independent of the star formation history -- these are
the black hole remnants of rapidly evolving high-mass stars
(initial mass larger than ca. 30 M$_{\sun}$, see Fig.~4). 
In the intermediate mass range of 
1 to 4 M$_{\sun}$ we see the expected differences: the largest
depletion of the MF at supersolar masses is found for the ``initial
burst'' case (i.e.~E and S0 galaxy stellar populations), the smallest
depletion for constant star formation rate (spiral and irregular
galaxy stellar populations), and the cosmic SFH case is lying
somewhere between, as the two principal galaxy classes roughly
contribute the same amount of stellar mass to the universal mean 
(see Read \& Trentham 2005). The (artificial) cut-off at 
$\log M \approx 1.4$
corresponds to the maximum final black-hole mass of ca.
25 M$_{\sun}$
originating from the initial-mass cut-off at 120 M$_{\sun}$.
As the high-mass end of the MF is entirely shaped by the 
very poorly constrained
final-initial mass function (Fig.~4), this part of
the distribution function should not be taken at face value.
Overall, from the viewpoint of a universal mass function 
of things, the present-day MF of stars and stellar remnants is not 
dramatically
different from the stellar IMF; for subsolar masses it is essentially
the same, for supersolar masses it is down in amplitude by roughly one order 
of magnitude.   

\begin{figure}[tb]
\plotone{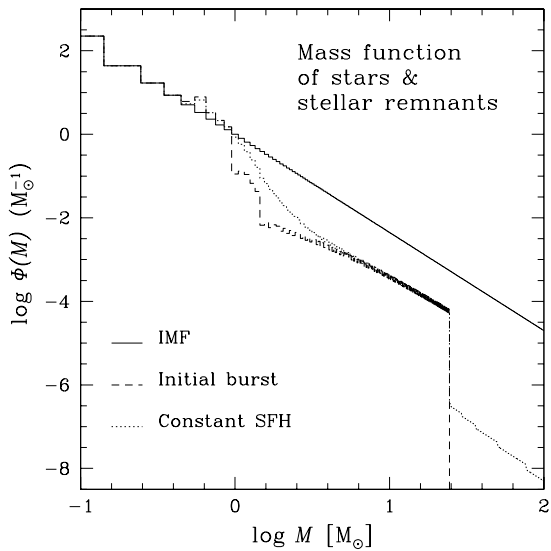}
\caption{Present-day mass functions of stars and stellar remnants
for a simple (co-eval) stellar population
of age 12 Gyr (``initial burst'', broken line), and a stellar population
with a constant star formation rate (dotted). The IMF (full line)
is shown for comparison.
The staircase character is a result of the binning (see text).
The normalization is arbitrary.  }
\end{figure}

\begin{figure}[tb]
\plotone{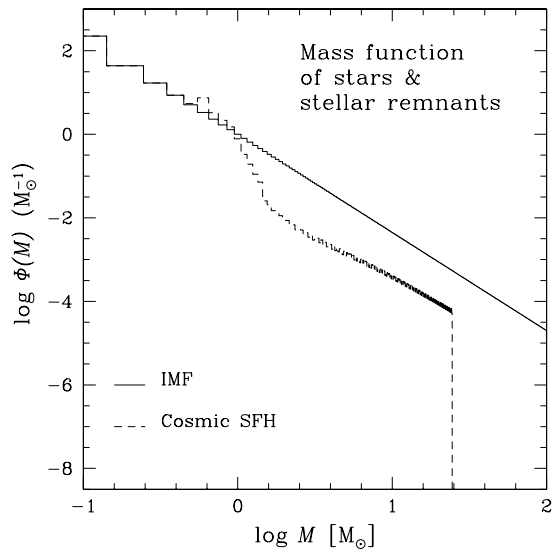}
\caption{Present-day mass function of stars and stellar remnants
for the assumed cosmic star formation history (broken line). 
Otherwise like Fig.~6}
\end{figure}

\subsubsection{Brown Dwarfs}
Brown dwarfs seem to cover a very narrow mass range, from the lower 
mass limit for hydrogen burning at $M$ = 0.08 M$_{\sun}$ downwards to about 
0.01 M$_{\sun}$, where the realm of giant planets is entered (see below).
The MF of brown dwarfs is much flatter than that of stars, even of low-mass 
stars. Luhmann et al.~(2000) and Bejar et al.~(2001),
studying brown dwarfs in young open clusters,
give $\alpha \approx 0.5-1$ for the usual powerlaw form 
$\phi(M) \propto M^{-\alpha}$. We adopt $\alpha = 0.8$ for objects with 
masses between 0.1 and 0.01 M$_{\sun}$ and attach this distribution 
continuously to the stellar one at 0.1 M$_{\sun}$. 

\subsubsection{Normalization}
The normalization of the stellar MF is straightforward. 
Read \& Trentham (2005) give a stellar contribution to the mean mass density
of the universe of $\Omega_* = 0.0028$. Hence we have
\begin{equation}
\Omega_* \rho_{\rm crit} = \hspace{-2mm}\int\limits_{
0.1 {\rm M}_{\sun}}^{25 {\rm M}_{\sun}} \hspace{-2mm}\phi_*(M) M dM
= A_* \hspace{-2mm}\int\limits_{0.1 {\rm M}_{\sun}}^{25 {\rm M}_{\sun}} 
\hspace{-2mm}M^{-x(M)} dM
\end{equation}
where $\phi_*(M)$ is the MF of stars and stellar remnants (from 0.1 to 25
M$_{\sun}$) as determined above for a cosmic star formation history 
(Fig.~7), and $x$ is the running (mass-dependent) power-law index of the 
same distribution. Brown dwarfs are not included; their contribution to the
total mass is negligible. In fact, for the integrated MF we could also
have neglected the stellar remnants. 
With $\rho_{\rm crit} = 1.36 \times 10^{11} {\rm M}_{\sun} {\rm Mpc}^{-3}$
this gives a normalization constant of $A_* = 7.6 \times 10^7 
{\rm Mpc}^{-3} {\rm M}_{\sun}^{-1}$. The normalized MF of stars and stellar
remnants and brown dwarfs appears in Figs.~9 to 11.

\subsection{Planets and Asteroids}
Inspite of the rapidly growing number of newly detected extrasolar planets,
our knowledge of the mass function of planetary and small bodies
has still to rely entirely on the census of our solar system.

\subsubsection{Solar System}
But even for the solar system we are far from having a complete census.
The recent detection of several Pluto-sized objects in the Kuiper belt
means that completeness of the solar system member census can be claimed
only to about a mass of $10^{-9} 
{\rm M}_{\sun}$, or a diamter of 1\,000 Km, which are roughly Pluto's
values. Taking data from the latest edition
of {\em Allan's Astrophysical Quantities}\/ (Cox 2000) for planets, moons,
and bright asteroids, and having turned the diameters of bright asteroids
and of some ``unweighted'' moons into masses by assuming a 
reasonable mean mass density of 
2 g cm$^{-3}$, we get a MF of solar system bodys as shown in Fig.~8
(indicated as crosses, giving numbers of objects per unit solar mass). 
This distribution can be
fitted by a straight line representing a rather shallow
power-law index of $-0.96$. 
As mentioned, below
$\approx 10^{-9} {\rm M}_{\sun}$ (Pluto's mass) 
this MF is bound to be incomplete. We construct a more reliable low-mass 
extension from faint asteroids, in the following way.

Ivezic et al. (2001) give a diameter
distribution function for 13\,000 asteroids found in the
{\em Sloan Digital Sky Survey}. The diameters had to be inferred from
magnitudes by assuming an albedo; Ivezic et al. (2001) assumed an albedo
of 0.14 for reddish (interpreted as mainly silicate, rocky) and an albedo 
of 0.04 for blueish (mainly carbonaceous) asteroids. The resulting diameter
($D$) distribution (see Ivezic et al., Fig. 25) can be fitted by a power law
of the form
\begin{equation}
N(D) dD \propto D^{-{\gamma}} dD\,\,,
\end{equation} 
where parameter $\gamma$, however, is different for different size regimes:
For asteroids larger than 5 km the distribution is very steep with 
$\gamma$ = 4, for smaller asteroids it is a bit more gentle with 
$\gamma$ = 2.3. This seems to be in good accord with other studies of
the size distribution of asteroids (see, e.g., Tedesco, Cellino \&
Zappala 2005;
Bagatin 2006). The size distribution is easily turned into a MF by again
assuming a mean mass density $\rho$ = 2 g cm$^{-3}$ for the asteroids.
With 
\begin{equation}
D = \left(\frac{6M}{\pi \rho}\right)^{\frac{1}{3}}
\end{equation}
we have
\begin{equation}
\phi (M) dM = \frac{dD}{dM} N(D) dM \propto M^{-\frac{(\gamma+2)}{3}} dM
\propto M^{-\gamma'} dM.
\end{equation}
The new power-law index $-\gamma'$ for the MF is now $-2$ for asteroids larger
than 5 km (= more massive than 6.6 $\times 10^{-17} {\rm M}_{\sun}$), 
and $-1.4$ for
smaller (less massive) bodies. The mass range covered extends from 
$10^{-20} {\rm M}_{\sun}$ (corresponding to asteroids of $D \approx$ 0.3 km)
up to $10^{-13} {\rm M}_{\sun}$ ($D \approx$ 100 km). This is the MF adopted
for our study, shown in Fig.~8. Lacking proper normalization, 
it is simply attached to the previously determined
MF for planets, moons, and bright asteroids by extrapolating
the steep, high-mass power law ($\gamma' = 2$) towards still 
higher masses, glueing the two distributions together
at $10^{-9} {\rm M}_{\sun}$, the assumed limit of completeness for the
planetary MF. By this we should more or less include the catalogued, brighter
(larger, more massive) asteroids missed by the {\em SDSS}-based work of 
Izevic et al. (2001; see Tedesco et al. 2005). Kuiper-belt (Trans-Neptunian)
objects seem to follow a distribution similar to the asteroids (e.g., 
Bagatin 2005, Fig. 2). 
By their exclusion, for lack of adequate data, we are likely
underestimating the MF amplitude for small bodies, but probably not more than
by a factor of 10, which is still not very relevant for our universal MF. 

\begin{figure}[tb]
\plotone{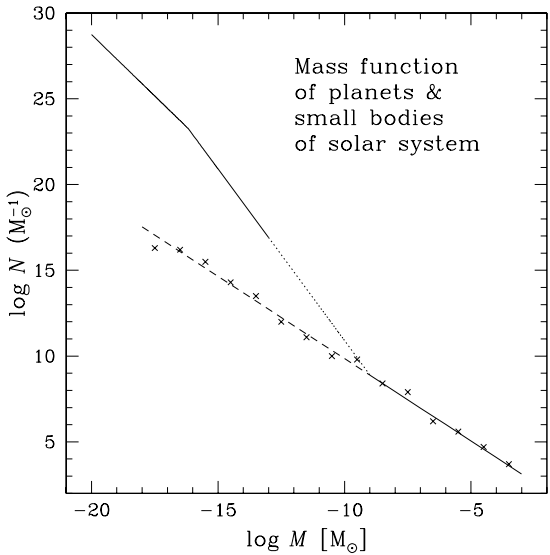}
\caption{Mass function of planets, satellites, and small bodies of the
solar system in numbers per unit solar mass. 
The crosses are differential counts
of solar system members based on catalogues -- deemed incomplete below a
mass of $10^{-9} {\rm M}_{\sun}$ (diameter around 1\,000 km). 
The straight line fitted through 
the crosses coresponds to a power-law index of $-0.96$. The full line at lower
masses with a knee around $10^{-16} {\rm M}_{\sun}$ (diameter $\approx$ 5 km)
is the diameter-inferred mass function of {\em SDSS}\/ asteroids. This 
distribution is extrapolated to higher masses (shown as dotted line) up to
$10^{-9} {\rm M}_{\sun}$, where it is attached to the planetary distribution.
}
\end{figure}

\subsubsection{Extrasolar Planets}
The exploration of extrasolar planets has only just begun. What we 
have learned so far is that the maximum planetary mass seems to be
around 10 Jupiter masses, or 1/100 of a solar mass -- which happens to coincide
roughly with the {\em minimum}\/ mass of brown dwarfs (see above). 
Most exoplanets 
detected so far have -- by necessity, given the detection techniques -- masses
in the narrow range
of 1 to 10 Jupiters (in fact, these are minimum masses, due to the unknown 
inclination of the planetary orbits). 
So we can use the (minimum) mass distribution of the 
presently known exoplantes (around 200, as of late 2006) to extend
the solar system MF towards 10 Jupiter masses. For this purpose we 
have collected data published in the web-based 
{\em Interactive Extrasolar Planets Catalog}\/ of Jean 
Schneider\footnote{
http://vo.obspm.fr/exoplanetes/encyclo/catalog.php} and glued the narrow
mass distribution function directly to the high-mass end of the solar system MF
(i.e. to Jupiter at 1/1000 of a solar
mass); this is the dotted curve in Fig.~9 indicated by ``Exoplanets''.  
Its exact form is irrelevant here, because it will change faster than we
could publish it!

\subsubsection{Normalization}
The universal abundance of planterary systems is of course completely
unknown. For solar-type (G-type) stars the frequency of planetary systems
is found 
to be in the 10 \% 
range (e.g., Quirrenbach 2005, p.39). But the detection
of extrasolar planets is still biased towards high masses and small periods;
so that freqency could be significantly higher. 
On the other hand, the majority of
stars are M dwarfs, for which planetary companions have yet to be 
discovered. The ubiquity of protoplanetary disks around low-mass stars
seems to suggest the ubiquity of planetary systems; but we do not yet know
whether this is generally true. Nor do we have a clue whether the 
abundance of planetary and small bodies scales with the mass of the star,
or whether the shape of the planetary MF is always the same, etc.
So, for want of better knowledge we make a very simple (solar-centric)
assumption: we assume that, on average in the universe, there is one solar 
system for every 10 solar masses 
(reproducing, of course, the 10 \% frequency for G-type stars).
Since, according to Read \& Trentham (2005), the mean stellar mass density 
of the universe is 
\begin{eqnarray}
\rho_* = \Omega_* \cdot \rho_{\rm crit} & = &
0.0028 \times 1.36 \ 10^{11} {\rm M}_{\sun} {\rm Mpc}^{-3}
\nonumber\\ & = &  
3.81 \times 10^{8} {\rm M}_{\sun} {\rm Mpc}^{-3}\,\,, 
\end{eqnarray}
we simply have to shift the solar system MF per unit solar mass shown in
Fig.~8 upwards (enhance the frequency) by a factor of 3.81 $\times 10^7$
to get a ``universal'' MF of substellar bodies per solar mass and 
per cubic Megaparsec. This is the distribution shown in Fig.~9.
It is apparent that the MFs of giant planets and brown dwarfs do not
smoothly interconnect, but there is also no reason why they should.
In fact, there seems to be a genuine frequency gap between the two types
of objects (e.g., Quirrenbach 2005, p.40). 
In any case, it should be borne in mind that this planet-star normalization
is probably the most unreliable normalization of our whole study. 

\begin{figure}[tb]
\plotone{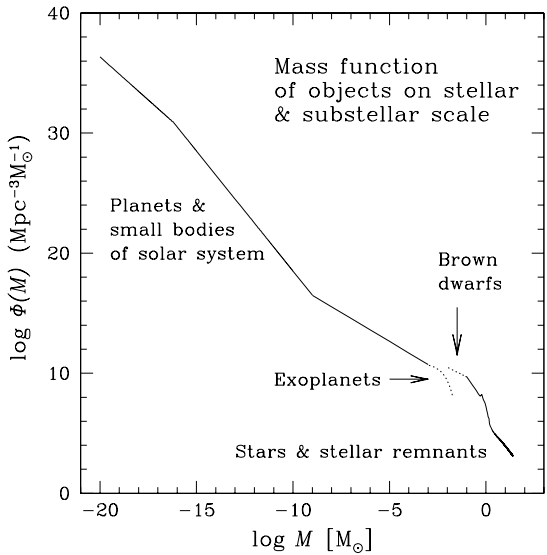}
\caption{The normalized mass function of solar system bodies and exoplanets,
along with the mass function of stars, stellar remnants and brown dwarfs.
Notice the possible discontinuity between giant planets and brown dwarfs.
}
\end{figure}

\section{PUTTING IT ALL TOGETHER: A UNIVERSAL MASS FUNCTION}
\subsection{Mass Function of Astronomical Objects}
We are now in a position that we can put together the MF for objects on 
superstellar scales (clusters and groups of galaxies, galaxies, star
clusters, and molecular clouds, Fig.~3) with the MF for objects
on the stellar and substellar scales (stars and stellar remnants,
brown dwarfs, planets, and asteroids, Fig.~9), to get an overall,
``universal'' MF that stretches from the most massive clusters of galaxies
-- the largest bound aggregates of matter in the universe at 
$M \approx 10^{16} {\rm M}_{\sun}$ -- all the way down to sub-kilometer-size
bodies of the solar system at $M \approx 10^{-20} {\rm M}_{\sun}$ 
(= 20 million tons), thus covering 36 (!) orders of magnitude in mass.
This is shown in Fig.~10 which constitutes the principal outcome of 
the present paper. It is gratifying that the two halves almost perfectly
connect to each other around one solar mass. Remember that this normalization
was achieved on the basis of the 
mean universal mass density carried by the stars, embodied in the galaxies
on the one hand, and comprised by the stars themselves on the other.
Before discussing the meaning of this and of the whole MF, we will 
in the following add 
one more piece of information: the MF of dark matter halos as derived from
theory and numerical simulations. 

\begin{figure}[tb]
\plotone{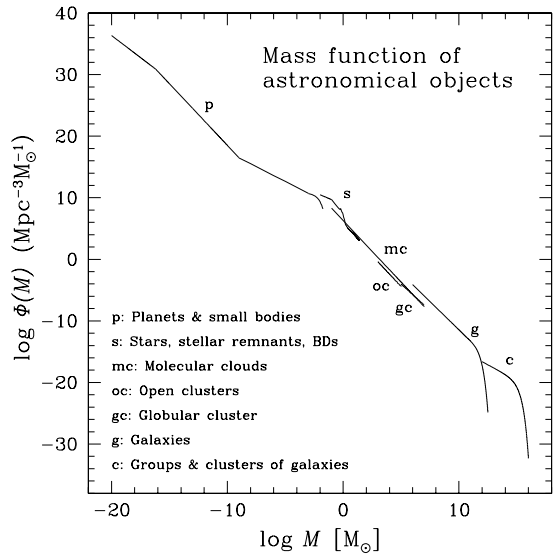}
\caption{The overall mass function of astronomical objects
from the most massive clusters of galaxies all the way down to 
sub-kilometer-size asteroids, covering 36 orders of magnitude in mass. 
The individual branches of objects are indicated. The figure is a 
combination of Figs.~3 and 9.}
\end{figure}

\subsection{Mass Function of Dark Matter Halos From Numerical Simulations}
A highly advanced theory of structure formation should one day 
be able to reproduce
the observed universal MF. Of course, we are very far from having reached
this point.
However, on larger scales the basic agent of structure formation is the
dissipationless gravitational clustering of matter, a process
which is fairly well
understood and is simulated,
with ever increasing numerical resolution, since more than 30 years. 
In the context of a cold dark matter (CDM)
scenario, starting with a Harrison-Zeldovich spectrum of primordial
density fluctuations, there is increasing 
fluctuation power with decreasing mass at recombination, 
getting asymptotically flat towards smaller masses (e.g., Longair 1998,
p.~316). This means that CDM structure is forming first on small scales
and is building up in a hierarchical way to ever larger scales. 
It is straightforward to show (e.g., Longair 1998, p.373, see also below), 
by using the formalism of 
Press \& Schechter (1974), that the flat low-mass end of the fluctuation 
spectrum at recombination results in a power-law MF
with slope $-2$ at small masses, i.e.
\begin{equation}
\phi(M)_{\rm DM} dM \propto M^{-2} dM\,\,.
\end{equation}
At the high-mass end, the MF rapidly turns down to zero, 
giving the functional form
a characteristic knee reminiscent of the Schechter luminosity 
function whose form was, in fact, motivated by the Press-Schechter 
result (Schechter 1976). The general form and low-mass slope of the DM
mass function has been 
confirmed and strengthened by more recent semi-analytic work
(Sheth \& Thormen (1999, see also Vale \& Ostriker 2004) and by numerous 
numerical simulations of the hierarchical collapse and clustering of DM halos
(e.g., Reed et al.~2003).

Recently, Diemand, Moore \& Stadel (2005a) performed very-high-resolution
numerical CDM simulations and found DM substructure right down to the 
theoretical limit at $M \approx 10^{-6} {\rm M}_{\sun}$, corresponding
to the mass of planet Earth (!). The MF of these ``microhalos''
turns out to have the 
same slope of $-2$ as the MF of more massive halos, at least down to
$M \approx 10^{-5} {\rm M}_{\sun}$ where the function starts to bend down,
depending on the nature of CDM (neutralinos or axions). Moreover,
in the mass range $10^{10}-10^{15} {\rm M}_{\sun}$
the normalization (absolute number of halos per solar mass and cubic Mpc)
of Diemand et al.'s simulation is only a factor of two different
from the one given by Reed et al.~(2003) 
when the latter MF is extrapolated down by 10 to 15 orders of
magnitude, with a forced slope of $-2$,
to reach the mass range of Diemand et al.~(2005a, see their Fig.~3).
 
For our study, in order
to have a theoretical counterpart of the MF of galaxies and
clusters of galaxies, we adopt a MF for DM halos of the form given in 
(24), i.e.~a slope of $-2$ for the entire mass range of 
$10^{-5}-10^{15} {\rm M}_{\sun}$ with the normalization of the Reed et al.
extrapolation given in Fig.~3 of Diemand et al.~(2005a). As mentioned, below 
$10^{-5} {\rm M}_{\sun}$ the MF is not known but will probably fall down
very rapidly. Above $10^{15} {\rm M}_{\sun}$ the MF bends down in the usual
``Schechter-Press'' manner (see e.g., Reed et al.~2003, Vale \& Ostriker 2004);
for this part of the MF we can directly rely on the cluster data.
The resulting, very simple theoretical MF for DM halos is shown in 
Fig.~11, along with all
other MFs shown already in Fig.~10. Obviously, there is good match
on the scale of galaxies and clusters of galaxies. This had also 
to be expected,
given the fact that, with respect to mass, DM is the major constituent
of these structures. In adition, the normalization of the simulated MF 
is of course not independent but is drawn from observed quantities, such as
the mean mass density of the universe and the density fluctuation amplitude 
encapsulated in $\sigma_8$. For galaxies, however, we note a difference: the
observed slope is flatter than the simulated slope. This means that the
simulations predict more substructure on galactic scales than is actually
observed, i.e.~most DM minihalos seem to lack a baryonic (dwarf galaxy)
counterpart. This discrepancy has manifested itself also
as ``satellite catastrophe'' (e.g., Moore at al.~1999). The relation 
of the mini/microhalos on still smaller scales to astronomical objects is
addressed below.

\begin{figure}[tb]
\plotone{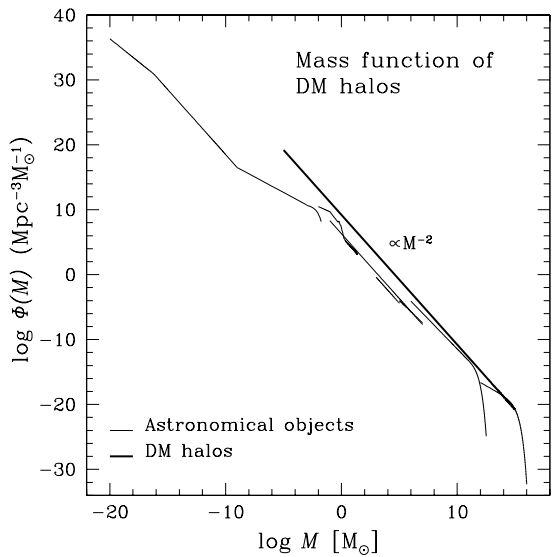}
\caption{The mass function of dark matter halos inferred from numerical
simulations, as compared to the observed mass functions of the various
classes of astronomical objects shown in Fig.~10.}
\end{figure}

\section{DISCUSSION AND CONCLUSIONS}
Here is first a recapitulation of our main findings:\\
(1) The mass functions of individual classes of astronomical objects
can be concatenated to build a nearly {\em continuous}, ``universal'' mass 
function 
that stretches from small asteroids all the way up to the richest 
clusters of galaxies, covering 36 orders of magnitude in mass.\\
(2) Most individual MFs follow, in part or entirely, a power law with index
$\alpha \approx -2$. A notable exception is the MF of planetary and 
small bodies
which is significantly flatter. As a consequence the combined MF
from stars to clusters of galaxies shows a ``universal'' slope of
$\alpha \approx -2$ as well.\\
(3) CDM halos from numerical simulations follow a nearly identical MF,
with $\alpha = -2$, holding even at the scale of stars and planets.\\
How should this be interpreted? What can be expected as a natural outcome
of structure formation? What cannot, and is in need of a special explanation?
Certainly the mass function(s) must somehow reflect
the physical processes involved with the
formation of these objects. 
Three different regimes of astronomical objects,
known to have radically different formation histories, 
can be distinguished:

(1) {\em Galaxies and supergalactic structures}\/ (from dwarf galaxies to rich
clusters of galaxies) were formed by gravitational clustering
in a {\em bottom-up}\/ manner. Their basic constituent 
is thought to be CDM, being clustered in underlying dark halos.
These halos (according to the most recent simulations) have a rich 
substructure down to the scale of stars and planets. The standard
formation scenario envisions the growth of 
primordial density fluctuations in CDM and the subsequent, hierarchical
gravitational instability and collapse of CDM halos from the smallest
(planetary) to the largest (galaxy cluster) scale. The $M^{-2}$ mass
distribution, which is roughly observed on the galactic and supergalactic 
scales and is implied by numerical simulations to hold 
for all scales, is in principle
well understood as a consequence of the power spectrum of primordial density
fluctuations, $P(k) \propto k^n$, and the ``transfer function'' 
describing the change of spectrum
due to the damping
of fluctuations in the pre-recombination epoch, $T(k) \propto k^{n'}$
in its most simple form. The resulting
power spectrum at recombination is $P_{\rm rec}(k) = 
P(k) \cdot T(k)^2 = k^{n + 2n'}$, and the low-mass end ($k \rightarrow \infty$)
of the mass function of DM halos at the present epoch becomes
(as inferred, e.g., from Efstathiou 1990, and Longair 1998):
\begin{equation}
\phi_{\rm DM} \propto M^{(n + 2n' - 9)/6}\,\,\,.
\end{equation}
With a Harrison-Zeldovich spectrum ($n = 1$) and for adiabatic CDM fluctuations
($n' = -2$ in the limit for large $k$), this turns into the familiar
$M^{-2}$ law. Clearly, the primordial spectral index is not as important
as the transfer function. So {\em the $M^{-2}$ behaviour 
can essentially be viewed
as a property of CDM}. Hot dark matter (HDM), e.g., would behave very
differently -- all fluctuations on large scales would get completely 
damped out and structure formation would (have to)
turn into a top-down scenario.

(2) {\em Stars and star clusters}\/ are believed to have 
formed by the gravoturbulent 
fragmentation of molecular clouds in a {\em top-down}\/ manner. 
It has been noted before that the mass functions of stars, star clusters,
and molecular cloud cores are quite similar. For stars we have the 
essentially still
valid Salpeter IMF with power-law slope $-2.35$, star clusters have mass 
functions with slope $\approx -2$ (see Sect.~2.3), and molecular gas
clouds show a fractal structure with fractal dimension $D \approx
2$, corresponding again to a MF of slope $-2$ (Testi \& Sargent 1998,
Elmegreen \& Falgarone 1996). In fact, a fractal structure with dimension
between 1.5 and 2.5 is generally found for the interstellar medium, 
galactic star fields, and even dwarf irregular galaxies (Elmegreen \&
Elmegreen 2001, Parodi \& Binggeli 2003). This MF universality
has been interpreted as direct
evidence for a universal formation mechanism for stars and star clusters in 
turbulent gas clouds (Elmegreen \& Efremov 1997). The origin of the 
scale-invariant clumpy structure of the gas itself is not known; it could 
result from gravitational fragmentation, collisional agglomeration, or 
turbulence (see references in Elmegreen \& Efremov 1997). Nor is it clear
what physical parameters determine the slope of the MF, why it is $-2$; 
the physics involved 
is exceedingly complex (see, e.g., Jappsen et al.~2005).

In the realm of stars and star clusters we should in principle also consider
the question of lifetimes, in addition to formation processes (this point was
brought to our attention by the referee of this paper). The lifetimes of
stars were 
explicitely dealt with in our calculation of the MF of an evolved stellar
population. Low-mass stars can get older than a Hubble time, anyway.  
The same is true for high-mass globular clusters.
However, {\em open clusters}\/ are known to have only a lifetime of a few
100 Myr (e.g., Piskunov et al. 2006). So one could indeed
wonder why they match
the extrapolated MF of stars so well (see Fig.~20). Apparently, their smaller
lifetimes (as compared to the bulk of stars) are
to a certain degree compensated
by a higher formation efficiency. This is in accord with the basic fact that
star formation efficiency is generally higher in high-density regions, 
which of course happen to be the loci of bound 
cluster formation (see Elmegreen 2006).  

In any case, it is a remarkable coincidence that, on the scale of stars and 
star clusters, the bottom-up CDM formation scenario should produce the same
slope, and even similar amplitude, as the top-down star formation
scenario.
The underlying physical processes, even if not exactly known,
are expected to be totally different. 
-- Or {\em is it conceivable that the DM substructure in galaxies has some
influence on the fragmentation of the gas, thereby imprinting 
on the cloud structure
the same ``universal'' mass distribution?}\/ This is difficult to imagine
but should not be excluded without further investigation. 
It was questioned whether microhalos could survive the 
tidal perturbations by other DM halos and, within the Galaxy, by stars and 
molecular clouds. But recent work (e.g., Goerdt et al. 2007) 
reaffirms that due to their
high density (cuspy structure) most microhalos should have survived.
If so, they might act on the gas in a manner that ``conserves'' the MF.
Although different processes to explain the stellar MF are usually
invoked (see above), minihalos (not microhalos) have already been causally
connected with old stars and globular clusters, if only 
in terms of their distribution and 
kinematics, and not yet their MF (Diemand, Madau \& Moore 2005b). 

(3) {\em Planets, moons, and small bodies}\/ were 
formed by the coagulation
of dust particles and the subsequent accretional growth of small
planetesimals
(giant planets additionally by the accretion of gas) in a {\em bottom-up}\/
manner. However, there are also {\em top-down}\/ processes on these scales.
For instance, asteroids formed (and as a population continuously 
evolve) by the 
collisional fragmentation of large planetesimals. The formation processes
for these objects are surely no less complex than for stars, and we cannot
even try to explain the MF on the scale of planets and small bodies in detail.
We also recall that the normalization of the planetary MF 
(their abundance {\em per cubic Mpc}) remains very uncertain,
simply because we know them only very nearby. However, one property
of the MF on these small scales is noteworthy: the general slope appears to be
significantly flatter than the ``universal'' $-2$. Again we can only
speculate on the reason for this difference. 
In the mass range from stars
to the largest structures in the universe, gravitation is certainly the
basic agent of structure formation, be it by bottom-up clustering 
or top-down fragmentation. On smaller scales, however,
gravitation becomes
less dominant and electromagnetic forces come into play; very small asteroids
(rocks),
e.g., are no longer held together by gravity but by van der Waals 
(hence electromagnetic) forces. So one could speculate
that the break in the slope of the MF around the mass of planets 
has something to do with the
{\em transition of the main force responsible for the constitution 
of objects from gravity to electromagnetic forces}. 

In conclusion, the title question of this paper can
certainly be answered by Yes. There is a universal mass function in the sense
that it is possible to put together a {\em continuous}\/ mass function 
``of the universe'': 
from asteroids and planets, over stars and stellar remnants, star clusters
and gas clouds, galaxies, all the way up to the richest custers of galaxies.
There is also a universal mass function in the sense that 
the slope of most individual
mass functions and that of the grand total is very similar:
$\phi(M) {\rm roughly} \propto M^{-2}$. We have no explanation for this
nearly universal slope for very different scales and processes. 
It is not even clear whether an explanation is needed, i.e.~whether it is
more than a chance coincidence.     

It would certainly be interesting to extend the MF of astronomical objects
from asteroids further down in mass to meter-size and submeter-size bodies
(which would have to include living things such as the reader!), 
to dust, molecules, atoms,
nuclei, and finally elementary particles. Such an -- even more speculative --
extension was not aimed at here.  

\acknowledgments
We thank Gustav Tammann, Helmut Jerjen, Pavel Kroupa, and Ralf Klessen 
for inspiration and fruitful discussions, Jay Gallagher for a crucial advice,
and the anonymous referee for interesting comments. Financial support by 
the Swiss National Science Foundation is gratefully acknowledged.

\end{document}